 \documentclass[pmlr,twocolumn,10pt]{jmlr} % W&CP article
\usepackage[T1,T2A]{fontenc}
\usepackage{multirow}
\usepackage{verbatim}
\usepackage{ulem}% \usepackage{geometry}
 \usepackage{rotating}% for sideways figures and tables
\usepackage{booktabs}
\usepackage[load-configurations=version-1]{siunitx} % newer version 
\theorembodyfont{\upshape}
\theoremheaderfont{\scshape}
\theorempostheader{:}
\theoremsep{\newline}

\jmlrvolume{LEAVE UNSET}
\jmlryear{2021}
\jmlrsubmitted{LEAVE UNSET}
\jmlrpublished{LEAVE UNSET}
\jmlrworkshop{Machine Learning for Health (ML4H) 2021} % W&CP title

\title[Comparative Study of Speech Analysis Methods to Predict Parkinson’s Disease]{Comparative Study of Speech Analysis Methods to Predict Parkinson’s Disease}

\author{
\Name{Adedolapo Aishat Toye} \Email{atoye@andrew.cmu.edu}\\
\addr Carnegie Mellon University Africa, Rwanda
\AND
\Name{Suryaprakash Kompalli} \Email{kompalli@cmu.edu}\\
\addr Carnegie Mellon University, USA \\
\addr INSOFE, India
}

\begin{document}

\maketitle

\begin{abstract}
One of the symptoms observed in the early stages of Parkinson's Disease (PD) is speech impairment. Speech disorders can be used to detect this disease before it degenerates. This work analyzes speech features and machine learning approaches to predict PD. Acoustic features such as shimmer and jitter variants, and Mel Frequency Cepstral Coefficients (MFCC) are extracted from speech signals. We use two datasets in this work: the MDVR-KCL and Italian Parkinson’s Voice and Speech database. To separate PD and non-PD speech signals, seven classification models were implemented: K-Nearest Neighbor, Decision Trees, Support Vector Machines, Naive Bayes, Logistic Regression, Gradient Boosting, Random Forests. Three feature sets were used for each of the models: (a) Acoustic features only, (b) All the acoustic features and MFCC, (c) Selected subset of features from acoustic features and MFCC. Using all the acoustic features and MFCC, together with SVM produced the highest performance with an accuracy of 98\% and F1-Score of 99\%. When compared with prior art, this shows a better performance. Our code and related documentation is available in a public domain repository\footnote{"https://github.com/aeesha-T/parkinsons\_prediction\_using\_speech"}
%This compares favorably with prior art. Our code and related documentation are available in a public domain.

\end{abstract}
\begin{keywords}
Parkinson’s disease, machine learning, speech signals, speech disorders, acoustic features, MFCC
\end{keywords}

\section{Introduction}
\label{sec:intro}
%Parkinson's disease (PD) is a neurological disorder that affects nerve cells in the brain leading to tremors, stiffness, or impaired balance \citet{disorders2020}. It is characterized with a certain part of the brain (that controls movement) being progressively damaged over time \citet{NHS2019, NIH2017}. The neurons in this part of the brain produce a chemical called dopamine. Dopamine is a neurotransmitter involved in the movement regulation in the brain \citet{Olguin2016, mishra2018}. When a patient has PD, these neurons become impaired or die, hence producing less dopamine. This leads to the impaired movements observed in PD patients \citet{NHS2019, NIH2017}.  
%While there is also no cure for PD \citet{NIH2017, parkinson2020}, treatments exist to provide symptomatic relief.

Parkinson's Disease (PD) is the second most common type of neurodegenerative disorder and it is generally observed in one out of every hundred people that are above 65 years of age \citet{Sriram2014, Sakar2013} and more than 10 million people are living with Parkinson's disease worldwide \citet{parkinson2020}. PD progresses slowly and the symptoms take years to develop. The main symptoms of this disease that are observed at later stages are tremors in the hand, rigidity (muscle stiffness), slowness of movement, and postural instability \citet{disorders2020}. In addition to these symptoms, PD is also known to cause disturbances in speech and reduced voice quality \citet{Sriram2014}. 

It has been reported that speech disorders symptomatic of PD can be detected as early as five years before a clinical diagnosis is made \citet{Zhang2016, Tsanas2012, Harel2004}. This lead time can be crucial to help predict PD before it degenerates. Speech impairments are characterized by dysphonia, which is the inability to produce vocal sounds, and dysarthria, which is the difficulty in pronouncing words \citet{Zhang2016}. These characteristics can be detected in the voice recordings of a patient. There is currently no specific test to diagnose PD \citet{disorders2020}. The current diagnosis is based on medical history, brain scans or blood tests to rule out other disorders. Treatments can be used to manage symptoms in the early stages of the disease, but this is not usually possible in the later stages. If automated analysis of speech can detect these disorders, it is likely that PD can be diagnosed at an early stage. An early diagnosis can be helpful in improving patient outcome. 

This research aims to create a repository of machine learning techniques to predict PD using raw speech data. To achieve this, we carry out speech analysis on the PD dataset described in \sectionref{sec:dataset} to extract speech features. Classifiers are then applied to these features to derive the best performing model to separate voice samples that are indicative of PD from samples that indicate normal speech. 

A variety of features from voice data have been used to predict this disease. Most studies tend to use Support Vector Machines, Random Forest, and Decision Tree algorithms \cite{Sriram2014, Zhang2016} to build classifier models. This study implements most of these algorithms. We use two datasets to perform a comprehensive comparison: The MDVR-KCL dataset \citet{mdvr2020}, and the Italian Parkinson's voice and speech database \cite{italian2017}. to the best of our knowledge, there have been no attempts to combine acoustic features and MFCC in predicting PD. The main contributions of this study are: (a) List algorithms and evaluation methods that may be useful for decision making using voice samples in clinical and semi-clinical scenarios, (b) Explore and compare the use of acoustic features (which include shimmer variants and jitter variants), MFCC and a combination of both in predicting PD, (c) Compare features and machine learning models for predicting PD, (d) Release a public domain repository of code. 

%\subsection{Parkinson Disease and Speech Disorder}
%Symptoms of PD progress slowly over the years and PD patients tend to experience adverse symptoms at later stages of the disease. The Hoehn and Yahr scale is often used by doctors to determine disease progression. This scale ranges from zero, (which represents no signs of PD) to five (which represents the most advanced form of PD). 
%There are two main categories of PD symptoms: motor and non-motor symptoms \citet{mishra2018,parkinson2020}. Motor symptoms include impaired movement, resting tremors, postural instability, and rigidity, while non-motor symptoms include depression, anxiety, irritability, apathy, constipation, loss of smell, sleep disturbances and learning and memory impairment \citet{NHS2019, mishra2018,fritsch2012, Tatiana2010}. Speech impairment is one of the significant symptoms of PD that occurs in most PD patients \citet{dashtipour2018}. It results from a combination of both the motor and non-motor symptoms experienced by the patients. The speech characteristics of PD include pitch fluctuations, voice tremor, roughness in the voice, reduction in speech volume, and oral festination \citet{dashtipour2018, nerurkar2017}. 
\section{Existing Work}
In 2020, \citet{Chethan2020} extracted 13 acoustic features (which include shimmer variants, jitter variants, pitch and Harmonics-to-Noise Ratio (HNR)) from the MDVR-KCL speech dataset \citet{mdvr2020} and used KNN classifier to predict PD. They achieved an accuracy of 85\% when all features were used in the classifier. However, when only optimized features were used, accuracies of 98\%, 94\% and 93\% were achieved for nine, seven, and four features, respectively. These features together with some additional acoustic features were also used by Berus et al \citet{Berus2018} to make predictions using the Parkinson’s dataset from the UCI Machine Learning Repository. Artificial Neural Networks, ANNs were used for the classification and Leave One Subject Out (LOSO) technique was used for the validation. This produced an accuracy of 86.47\%.

Other spectral features have also been explored, such as Mel Frequency Cepstral Coefficients (MFCC), and Intrinsic Mode Function (IMF). For instance, \citet{Jeancolas2017} proposed the use of Mel-Frequency Cepstral Coefficients (MFCC) combined with Gaussian Mixture Models (GMM) to detect early stages of PD. An accuracy of 91.4\% and 79.5\% were obtained for men and women, respectively. \citet{Tripathia2020} also proposed a novel approach for predicting PD, using the Italian Parkinson’s Voice and Speech database \citet{italian2017}. This approach used the Empirical Mode Decomposition technique to transform the speech data into its Intrinsic Mode Functions (IMF), which were then used to train a 1D-CNN model. Accuracy figures of 87\% and 81\% were reported for correct prediction of PD (Parkinsons Disease) and HC (Healthy Control) subjects respectively.

\section{Dataset}
\label{sec:dataset}
%\subsection{Feature Extraction}
Two datasets have been used in this study:  Italian Parkinson's Voice and Speech Database \citet{italian2017} and the mobile device voice recordings at King's College London (MDVR-KCL) \citet{mdvr2020}. 

The Italian Parkinson's Voice and Speech Database \citet{italian2017} consists of 15 healthy people within the ages of 19-29 years, 22 healthy people within the ages of 60-77 years, and 28 PD patients within the ages of 40-80 years.
%\sout{How many total voice recordings are there?}

The PD patients were confirmed by specialists to have a severity of less than 4 on the Hoehn and Yahr scale, except for a patient with stage 4 and another with stage 5. All participants are from Italy. The participants took part in reading exercises and pronunciation of vowels in an echo-free room while standing 15cm to 25cm away from the microphone. Several voice samples are present for each participant.  In this study, we used a total of 495 recordings of vowel pronunciations from the dataset. 

The MDVR-KCL dataset consists of mobile device voice recordings of 37 participants, carried out at King's College London in September, 2017 \cite{mdvr2020}. The dataset consists of voice recordings from 16 PD patients and 21 healthy controls subjects. A total of 37 recordings of participants reading the same paragraph of text is available. The recordings are stored in .wav format with each participant data being labelled with the Hoehn \& Yahr (H\&Y) score, the UPDRS II part 5 and the UPDRS III part 18 scale rating, indicating the stage of the disease. These ratings were based on an expert assessment.

%\textbf{
The duration of each recording is approximately 150 seconds. In this study, we divided each recording into smaller segments by splitting at locations where the speaker was silent for half a second or more. This generated approximately 22 segments per recording, to get a total of 808 samples that were used for further analysis. 

%\sout{How many total voice recordings are there?}

%The recordings were made using an Android recording application that runs as a stand-alone background service on a Motorola Moto G4 smartphone. All recordings were made such that the subject and the microphone were within the reverberation radius, i.e., the distance from the sound source at which the sound still persists. Each recording consists of the participant reading out a portion of text.

\section{Methodology}
\label{sec:methodology}
\subsection{Feature Extraction}
The voice recordings underwent a feature extraction process in which all the needed features were extracted from the voice speech data. The features extracted are the acoustic features and MFCC features. For this study, the acoustic features extracted for each speech signal include:  Jitter (absolute),  Jitter (relative),  Jitter (rap),  Jitter (PPQ5),  Shimmer(db), Shimmer(relative),  Shimmer(APQ3), Shimmer(APQ5), Fundamental Frequency, Harmonics-to-Noise Ratio (HNR), and Pitch. These features have been used previously in classification of PD patients \citet{Chethan2020, Berus2018}.

For each speech signal, we built a 2D matrix (F * 13) of MFCC values where F represents the number of frames and 13 is the number of MFCC coefficients extracted \cite{Jeancolas2017, Benba2015}. To convert this 2D matrix to a single vector, the mean of all frames was calculated to obtain a vector of 13 values. These are then normalized before being used to train the classifiers. The acoustic and MFCC features are well known in literature and for reference, their computation is included in appendix \ref{apd:features}. 

%explained in \sectionref{sec:speech}.
To pre-process the data, outlier analysis and scaling with Min-Max was performed. The dataset was split into 70\% training and 30\% validation set. 

\subsection{Classification Methods}
The classification algorithms used in this study are: KNN, Decision Trees, Support Vector Machines(SVM), Naïve Bayes, Logistic Regression, Gradient Boosting and Random Forest. These algorithms were all built using the scikit-learn library. GridSearch was used as the hyperparameter tuning method to select the optimal parameters for each of the models except Naïve Bayes.

\section{Experiment Details}
Each of the models was evaluated using repeated k-fold cross validation, where k is six. The performance was then recorded using accuracy, recall, specificity, precision, and F1-score. This was done ten times to verify the stability of the models. 

Three experiments were carried out using the features discussed in \sectionref{sec:methodology}. The first experiment involved using only the acoustic features, while the second combined all the acoustic features and the 13 MFCC values. The third experiment used a set of selected features from a combination of both acoustic and MFCC features. 
%The SelectKBest function in the scikit-learn library was used to carry out the feature selection process. 
We used the ANOVA f-statistic to select the best features. This verifies if the means of two or more groups of the sample data differ significantly from each other. This analysis provides a score to each feature, whereby a feature with the highest score is regarded as the most important feature. This process resulted in the selection of the following features for the Italian Parkinson's Voice and Speech Database: HNR, shimmer (relative), shimmer(absolute), shimmer(APQ3), shimmer(APQ5), and the MFCCs.

%\sout{ This helps to indicate the features that are most likely not important to the classification problem because they are independent of the target labels. The selected features varies for each experiment. They include a subset of the following: mean fundamental frequency, fundamental frequency (standard deviation), HNR, jitter (absolute), jitter(relative), jitter(rap), jitter(PPQ5), shimmer (relative), shimmer(absolute), shimmer(APQ3), shimmer(APQ5), and the 13 MFCCs. } This text is very confusing. "The selected features varies for each experiment." makes no sense. If each experiment has different features, how can some features be better? Rephrase. Possible rephrasing: {A: I meant to say, the selected features when training with different datasets differs }
%{\bf List out selected features here.}

\section{Results and Conclusions}
%The models were evaluated on the Italian dataset using the acoustic features. This consists of 11 features. We used Repeated CV for evaluation.

%\sout{ The language used and words have to be simple. Avoid using complex sentences}

\tableref{tab:italian_performance} shows the results of our work using the Italian Parkinson's Voice and Speech dataset. In the first experiment which used 11 acoustic features, %\sout{ SVM is observed to have } 
SVM has the best performance with accuracy of 87.2\%, specificity of 83.3\%, 87\% precision, 90\% sensitivity and f1-score of 88.6\%. Logistic Regression beats SVM in sensitivity: 91.1\%. 

In the second experiment all acoustic and MFCC features are combined into a feature vector of 24 attributes. SVM has the overall best performance with accuracy  98.3\%, specificity 98.7\%, sensitivity 98\%, precision 98.9\% and f1-score 98.4\%. 

In the third experiment, top ten features were selected from the 24 features. SVM performed better than other models with accuracy 98.9\%, specificity 99\%,  sensitivity 98.8\%, precision 99.2\% and f1-score 99\%.

\begin{table*}
\floatconts
  {tab:italian_performance}
  {\caption{Average performance of each model on Parkinson's Voice and Speech Database.}}%
  {
    \setlength{\tabcolsep}{3pt}
    \begin{tabular}{|c|l|l|l|l|l|l|l|l|} 
    \hline
    Features & Metrics & KNN & DT &SVM & NB & LR & GB & RF\\ \hline
    \multirow{5}{4em}{11 Acoustic features} & Accuracy & 81.6 & 77 & 87.2 & 65.6 & 80 & 84.8 & 83.8\\ 
    & Specificity & 80.9 & 75.8 & 83.3 & 34.9 & 66.4 & 83.1 & 81.6\\ 
    & Recall & 82.2 & 78 & 90.4 & 90.2 & 91.1 & 86.2 & 85.7\\
    & Precision & 84.3 & 80.2 & 87 & 63.4 & 77.2 & 86 & 85\\
    & F1 score & 83 & 78.9 & 88.6 & 74.4 & 83.4 & 86.2 & 85.4\\ \hline
    
    \multirow{5}{4em}{13 MFCC and 11 acoustic features} & Accuracy & 96.7 & 86.6 & 98.3 & 74.7 & 95.6 & 97.3 & 93.7\\ 
    & Specificity & 98.2 & 84.9 & 98.7 & 51.3 & 96 & 97.6 & 91\\ 
    & Recall & 95.4 & 88 & 98 & 93 & 95 & 97 & 96\\
    & Precision & 98.5 & 87.9 & 98.9 & 70.7 & 96.7 & 98.1 & 93\\
    & F1 score & 97 & 87.9 & 98.4 & 80 & 96 & 97.6 & 94.3\\ \hline
  
    \multirow{5}{4em}{10 selected features} & Accuracy & 97 & 86.2 &  98.9  & 74.2 & 93.2 & 96.7 & 94.3\\ 
    & Specificity & 98.7 & 84.6 & 99 & 51 & 91 & 96.8 & 92.8\\ 
    & Recall & 95.7 & 87.4 & 98.8  & 92.8 & 95 & 96.8 & 95.7\\
    & Precision & 98.9 & 87.7 & 99.2 & 70.3 & 93 & 97.3 & 94.2\\
    & F1 score & 97.2 & 87.5 & 99 & 79.9 & 93.9 & 97 & 94.9\\ \hline
  
    \end{tabular}
    }
\end{table*}

\begin{table*}[htbp]
\floatconts
  {tab:mdvr-kcl_performance}
  {\caption{Average performance of each model on the MDVR-KCL Dataset}}%
  {
    \setlength{\tabcolsep}{3pt}
    \begin{tabular}{|c|l|l|l|l|l|l|l|l|} 
    \hline
    Features & Metrics & KNN & DT &SVM & NB & LR & GB & RF\\ \hline
    \multirow{5}{4em}{11 Acoustic features} & Accuracy & 77.2 & 69.2 & 77.7 & 67.8 & 74.7 & 75.6 & 78.5 \\ 
    & Specificity & 87 & 72.6 & 89.7 & 77.2 & 87.3 & 95.9 & 87.3\\ 
    & Recall & 63.7 & 64.7 & 61.1 & 54.7 & 57.4 & 47.5 & 66.3\\
    & Precision & 78.3 & 63.1 & 81.3 & 64 & 76.6 & 89.4 & 79.2\\
    & F1 score & 70 & 63.6 & 69.6 & 58.3 & 65.4 & 61.8 & 71.9\\ \hline
    
    \multirow{5}{4em}{13 MFCC and 11 acoustic features} & Accuracy & 90.9 & 80.3 & 86.3 & 73.2 & 80.1 & 75.4 & 86.9\\ 
    & Specificity & 95.2 & 82.2 & 95.3 & 77.9 & 88 & 95.3 & 93.2\\ 
    & Recall & 85.2 & 77.7 & 73.8 & 66.7 & 69.3 & 48 & 78.5\\
    & Precision & 92.8 & 76.1 & 92 & 68.8 & 80.6 & 88.4 & 89.2\\
    & F1 score & 88.7 & 76.7 & 81.7 & 67.4 & 74.3 & 62 & 83.4 \\ \hline
  
    \multirow{5}{4em}{10 selected features} & Accuracy & 85 & 76.7 & 79.9 & 72.1 & 75.4 & 75.4 & 84\\ 
    & Specificity & 91.6 & 80 & 89.9 & 78.3 & 84.1 & 95.4 & 90.7\\ 
    & Recall & 76.1 & 72.2 & 66.2 & 63.4 & 63.6 & 47.7 & 74.9\\
    & Precision & 86.7 & 72.2 & 82.7 & 68 & 74.4 & 88.2 & 85.2\\
    & F1 score & 80.9 & 71.9 & 73.2 & 65.4 & 68.4 & 61.7 & 79.5\\ \hline
  
    \end{tabular}
    }
\end{table*}

%\sout{We observed that } 
While SVM with 10 selected features gives the best result, all methods do not perform better while using the 10 selected features. For example, Linear Regression and Gradient Boosting give better accuracy when using all 24 features (11 acoustic + 13 MFCC). 

\tableref{tab:mdvr-kcl_performance} displays the results of our work using the MDVR-KCL dataset. KNN using all 24 features gives the best results. 
\tableref{tab:prior_work} shows a comparison of our work with prior art. The work by \cite{Chethan2020} mentions the use of genetic algorithms for feature extraction to obtain an accuracy of 99.2\% on the MDVR-KCL dataset. Details of the hyper parameters used are not available in the published work, and we have been unable to reproduce their results. %\textbf{ 
Another aspect related to MDVR-KCL dataset is that we were using a simplistic method to detect silent spaces and segment the approximately 150 second long audio files into segments from which features were extracted. The segmentation is not consistent across different speakers, and can contribute to lack of accuracy. We are exploring alternative methods of speech signal segmentation to improve the overall prediction accuracy. 

In this study, we have been able to identify the optimal speech features and best performing machine learning technique that can be used to predict PD. We have provided a publicly accessible codebase with the implementation of our work. This improvement in performance and publicly available implementations can help in building applications for better diagnosis of PD. 

One of the limitations of research into voice analysis for PD is the lack of abundant large sample size voice recordings. %\sout{
The MDVR-KCL dataset has 37 recordings from 37 individuals. The Italian Parkinsons' Voice and Speech Dataset has a total of 495 recordings of vowels pronunciation from 65 individuals. The models we have chosen are suitable for these dataset sizes. Some of the more recent methods like LSTMs require larger datasets. Future research can include data collection or data augmentation of voice recordings for PD research.  Investigating the variation observed while training on different datasets would also be important. Other speech features such as linear prediction cepstral coefficients (LPC), and discrete wavelet transform (DWT) \cite{Alim2018}  can be explored.

%There is still some room for improvement in this field. Other speech features such as linear prediction cepstral coefficients (LPC), and discrete wavelet transform (DWT), can be explored to determine its importance in detecting voice disorders. Also, additional research can be done to test these models on real-life unknown data to verify its performance in a real-life scenario. 

\begin{table*}[htbp]
\floatconts
%\footnotesize
   {tab:prior_work}
   {\caption{Comparison of our work with prior art}}
   {%\begin{center}
\begin{tabular}{ | p{3cm} | p{3cm} | p{4cm} | p{2cm} |}
\hline
\textbf{Reference} & \textbf{Dataset Used} & \textbf{Method} & \textbf{Accuracy}\\ \hline
\citet{Chethan2020} &  MDVR-KCL for PD \cite{mdvr2020} & Hand crafted features/ KNN & 98\%\\ \hline
\citet{Tripathia2020} & Italian Parkinson's Voice and Speech database \citet{italian2017} & IMF signals/1D-CNN & 84.6\%\\ \hline
\citet{kurada2020} & MDVR-KCL for PD & VGGish feature extraction/KNN & 87\% \\ \hline
\multirow{7}{3em}{Our Work} & Italian Parkinson's Voice and Speech Database & 11 acoustic features + 13 MFCC & \\
& & SVM & \bf 98.9 \%\\ 
& & KNN & 97 \%\\ 
& & DT & 86.2 \%\\ 
& & NB & 74.2 \%\\ 
& & LR & 93.2 \%\\ 
& & GB & 96.7 \%\\ 
& & RF & 94.3 \%\\ \hline

\multirow{7}{3em}{Our Work} & MDVR-KCL & 11 acoustic features + 13 MFCC & \\
& & SVM & 86.3 \%\\ 
& & KNN & \bf 90.9 \%\\ 
& & DT & 80.3 \%\\ 
& & NB & 73.2 \%\\ 
& & LR & 80.1 \%\\ 
& & GB & 75.4 \%\\ 
& & RF & 86.9\%\\ \hline

\end{tabular}
}
%\end{center}
\end{table*}

%\nocite{*}
\bibliography{jmlr-sample}

\begin{thebibliography}{24}
\providecommand{\natexlab}[1]{#1}
\providecommand{\url}[1]{\texttt{#1}}
\expandafter\ifx\csname urlstyle\endcsname\relax
  \providecommand{\doi}[1]{doi: #1}\else
  \providecommand{\doi}{doi: \begingroup \urlstyle{rm}\Url}\fi

\bibitem[Alim and Rashid(2018)]{Alim2018}
S.~Alim and N.~Rashid.
\newblock Some commonly used speech feature extraction algorithms.
\newblock From Natural to Artificial Intelligence-Algorithms and Applications,
  2018.

\bibitem[Benba et~al.(2015)Benba, Jilbab, Hammouch, and Sandabad]{Benba2015}
A.~Benba, A.~Jilbab, A.~Hammouch, and S.~Sandabad.
\newblock Voiceprints analysis using mfcc and svm for detecting patients with
  parkinson's disease.
\newblock In \emph{2015 International Conference on Electrical and Information
  Technologies (ICEIT)}, pages 300--304, 2015.
\newblock \doi{10.1109/EITech.2015.7163000}.

\bibitem[Berus et~al.(2018)Berus, Klancnik, Brezocnik, and Ficko]{Berus2018}
L.~Berus, S.~Klancnik, M.~Brezocnik, and M.~Ficko.
\newblock Classifying parkinson's disease based on acoustic measures using
  artificial neural networks.
\newblock \emph{Sensors (Basel)}, 19\penalty0 (1), 2018.
\newblock ISSN 2321-9653.

\bibitem[Chavan and Munot(2016)]{chavan_2016}
A.~S. Chavan and S.~S Munot.
\newblock Effect of pre-processing along with mfcc parameters in speech
  recognition.
\newblock \emph{International Journal of Engineering Development and Research
  (IJEDR)}, 4:\penalty0 563--566, 2016.

\bibitem[Chethan et~al.(2020)Chethan, Abhishek, Divitha, Nagaraju, and
  Aditya]{Chethan2020}
C.~Chethan, M.S. Abhishek, D.~Divitha, T.R Nagaraju, and C.~R. Aditya.
\newblock Diagnosis of parkinson disorder through speech data.
\newblock \emph{International Journal for Research in Applied Science and
  Engineering Technology (IJRASET)}, 8\penalty0 (4), 2020.
\newblock ISSN 2321-9653.

\bibitem[Dimauro and Girardi(2017)]{italian2017}
G.~Dimauro and F.~Girardi.
\newblock Italian parkinsons voice and speech, 2017.

\bibitem[Fayek(2016)]{fayek2016}
Haytham~M. Fayek.
\newblock Speech processing for machine learning: Filter banks, mel-frequency
  cepstral coefficients (mfccs) and what's in-between, 2016.

\bibitem[Fernandes et~al.(2018)Fernandes, Teixeira, Guedes, Junior, and
  Teixeira]{Joana2018}
J.~Fernandes, F.~Teixeira, V.~Guedes, A.~Junior, and J.~Teixeira.
\newblock Harmonic to noise ratio measurement - selection of window and length.
\newblock \emph{Procedia Computer Science}, 138:\penalty0 280--285, 2018.
\newblock ISSN 1877-0509.

\bibitem[Ferrand(2002)]{ferrand2002}
C.~T. Ferrand.
\newblock Harmonics-to-noise ratio.
\newblock \emph{Journal of Voice}, 16\penalty0 (4):\penalty0 480--487, 2002.

\bibitem[Harel et~al.(2004)Harel, Cannizzaro, and Snyder]{Harel2004}
B.~Harel, M.~Cannizzaro, and P.~J. Snyder.
\newblock Variability in fundamental frequency during speech in prodromal and
  incipient parkinsons disease: A longitudinal case study.
\newblock \emph{Brain and Cognition}, 56\penalty0 (1):\penalty0 24--29, 2004.
\newblock ISSN 0278-2626.

\bibitem[Jaeger et~al.(2020)Jaeger, Trivedi, and Stadtschnitzer]{mdvr2020}
H.~Jaeger, D.~Trivedi, and M.~Stadtschnitzer.
\newblock Mobile device voice recordings at king's college london (mdvr-kcl)
  from both early and advanced parkinson's disease patients and healthy
  controls, 2020.

\bibitem[Jeancolas et~al.(2017)Jeancolas, Benali, and et~al]{Jeancolas2017}
L.~Jeancolas, H.~Benali, and B.~Benkelfat et~al.
\newblock Automatic detection of early stages of parkinson disease through
  acoustic voice analysis with mel-frequency cepstral coefficients.
\newblock \emph{ATSIP 2017: 3rd International Conference on Advanced
  Technologies for Signal and Image Processing}, 2017.

\bibitem[Jiménez-Hernández(2016)]{mario2016}
M.~Jiménez-Hernández.
\newblock A tutorial to extract the pitch in speech signals using
  autocorrelation.
\newblock \emph{Open Journal of Technology and Engineering Disciplines
  (OJTED)}, 2\penalty0 (1):\penalty0 01--10, 2016.

\bibitem[Kurada and Kurada(2020)]{kurada2020}
S.~Kurada and A.~Kurada.
\newblock Poster: Vggish embeddings based audio classifiers to improve
  parkinson’s disease diagnosis.
\newblock \emph{IEEE/ACM International Conference on Connected Health:
  Applications, Systems and Engineering Technologies (CHASE)}, 2020.

\bibitem[NINDS(2020)]{disorders2020}
NINDS.
\newblock Parkinsons disease: hope through research, August 2020.

\bibitem[ParkinsonsFoundation(2020)]{parkinson2020}
ParkinsonsFoundation.
\newblock Understanding parkinson, August 2020.

\bibitem[Sakar et~al.(2013)Sakar, C.Sakar, Isenkul, and Sertbas]{Sakar2013}
B.~Sakar, C.Sakar, M.~Isenkul, and A.~Sertbas.
\newblock Collection and analysis of a parkinson speech dataset with multiple
  types of sound recordings.
\newblock \emph{IEEE Journal of Biomedical and Health Informatics}, 17\penalty0
  (4), 2013.

\bibitem[Singh and Rani(2014)]{parwinder2014}
P.~P. Singh and P.~Rani.
\newblock An approach to extract feature using mfcc.
\newblock \emph{IOSR Journal of Engineering (IOSRJEN)}, 4:\penalty0 21--25,
  2014.

\bibitem[Sreenivasa and Manjunath(2017)]{rao_2017}
R.~K. Sreenivasa and K.~E. Manjunath.
\newblock Speech recognition using articulatory and excitation source features.
\newblock \emph{SpringerBriefs in Electrical and Computer Engineering}, 2017.

\bibitem[Sriram et~al.(2014)Sriram, Rao, Narayana, and Kaladhar]{Sriram2014}
T.~Sriram, M.~Rao, G.~Narayana, and D.~Kaladhar.
\newblock Diagnosis of parkinson disease using machine learning and data mining
  systems from voice dataset.
\newblock \emph{Proceedings of the 3rd International Conference on Frontiers of
  Intelligent Computing: Theory and Applications (FICTA)}, 327, 2014.

\bibitem[Tripathia and Kopparapua(2020)]{Tripathia2020}
A.~Tripathia and S.~K. Kopparapua.
\newblock Cnn based parkinsons disease assessment using empirical mode
  decomposition.
\newblock \emph{Proceedings of the CIKM 2020 Workshops}, 2020.

\bibitem[Tsanas et~al.(2012)Tsanas, Little, McSharry, Spielman, and
  Ramig]{Tsanas2012}
A.~Tsanas, M.~Little, P.~McSharry, J.~Spielman, and L.~Ramig.
\newblock Novel speech signal processing algorithms for high-accuracy
  classification of parkinson's disease.
\newblock \emph{IEEE Transactions on Biomedical Engineering}, 59\penalty0
  (5):\penalty0 1264 -- 1271, 2012.

\bibitem[Xuanpei et~al.(2001)Xuanpei, Hsiao-Wuen, and Alex]{huang_2001}
H.~Xuanpei, H.~Hsiao-Wuen, and A.~Alex.
\newblock \emph{Spoken Language Processing: A guide to theory, algorithm, and
  system development}.
\newblock Prentice Hall PTR, 2001.

\bibitem[Zhang et~al.(2016)Zhang, Yang, and et~al]{Zhang2016}
H.~Zhang, L.~Yang, and Y.~Liu et~al.
\newblock Classification of parkinson disease utilizing multi-edit
  nearest-neighbour and ensemble learning algorithms with speech samples.
\newblock \emph{BioMedical Engineering OnLine}, 15\penalty0 (122), 2016.

\end{thebibliography}

\appendix

\section{Acoustic and MFCC Features}\label{apd:features}
\subsection{Acoustic Features}

{\bf Jitter (absolute)} 

This is the average absolute difference between two consecutive periods.

 \[\frac{1}{N-1} \sum _{i=1}^{N-1} \lvert T_{i} - T_{i+1} \rvert \] 
 
where: \\
N = number of periods \\
T = period
 
{\bf Jitter (relative)}

This is the average absolute difference between two consecutive periods divided by the average period.

\[\frac{\frac{1}{N-1} \displaystyle \sum _{i=1}^{N-1} \lvert T_{i} - T_{i+1} \rvert}{\frac{1}{N} \displaystyle \sum _{i=1}^{N} T_{i}} \times 100 \] 

where: \\
N = number of periods \\
T = period

{\bf Jitter (rap)}
This is the average absolute difference between a period and the average of it and its two neighbours, divided by the average period in percentage.

\[\frac{\frac{1}{N-1}\displaystyle \sum _{i=1}^{N-1} \lvert T_{i} - (\frac{1}{3} \sum _{n=i-1}^{i+1}T_{n}) \rvert}{\frac{1}{N} \displaystyle \sum _{i=1}^{N} T_{i}} \times 100\] \\

where: \\
N = number of periods \\
T = period

{\bf Jitter (PPQ5)}
This is the average absolute difference between a period and the average of it and its four closest neighbours, divided by the average period in percentage.

\[\frac{\frac{1}{N-3}\displaystyle \sum _{i=2}^{N-2} \lvert T_{i} - (\frac{1}{5} \sum _{n=i-2}^{i+2}T_{n}) \rvert}{\frac{1}{N} \displaystyle \sum _{i=1}^{N} T_{i}} \times 100\] \\

where: \\
N = number of periods \\
T = period

{\bf Shimmer(db)}\\
This is the average absolute of log of the ratio between the amplitudes of consecutive periods, multiplied by 20. It is measured in decibels.

 \[\frac{1}{N-1} \sum _{i=1}^{N-1} \lvert  20 \times log(\frac{A _{i+1}}{A_{i}}) \rvert \] \\

{\bf Shimmer(relative)}\\
This is the average absolute difference between the amplitudes of two consecutive periods divided by the average amplitude.

\[\frac{\frac{1}{N-1} \sum _{i=1}^{N-1} \lvert A_{i} - A_{i+1} \rvert}{\frac{1}{N} \displaystyle \sum _{i=1}^{N} A_{i}} \times 100\]

{\bf Shimmer(APQ3)}\\
This is the average absolute difference between the amplitude of a period and the average of the amplitudes of it and its neighbours, divided by the average amplitude.

\[\frac{\frac{1}{N-1}\displaystyle \sum _{i=1}^{N-1} \lvert A_{i} - (\frac{1}{3} \sum _{n=i-1}^{i+1}A_{n}) \rvert}{\frac{1}{N} \displaystyle \sum _{i=1}^{N} A_{i}} \times 100\] \\

{\bf Shimmer(APQ5)}\\
This is the average absolute difference between the amplitude of a period and the average of the amplitudes of its and its four neighbours, divided by the average amplitude.

\[\frac{\frac{1}{N-3}\displaystyle \sum _{i=2}^{N-2} \lvert A_{i} - (\frac{1}{5} \sum _{n=i-2}^{i+2}A_{n}) \rvert}{\frac{1}{N} \displaystyle \sum _{i=1}^{N} A_{i}} \times 100\] \\

{\bf Fundamental Frequency}\\
Frequency is the average number of oscillations/cycles per second. The fundamental frequency is known as the lowest frequency of a periodic waveform. It is measured in Hertz.
\\

%\subsection{Harmonics}
%Harmonics are multiples of the fundamental frequency. They have lower amplitude than the F0. For example, if the fundamental frequency is 100Hz, the second harmonic will vibrate twice as fast as 100Hz, that is 200Hz. The third harmonic will vibrate at  300Hz.

{\bf Harmonics-to-Noise Ratio (HNR)}\\
HNR quantifies the relationship between the periodic and non-periodic components of the speech signal by calculating the ratio between the two components. This feature indicates the voiced components of the speech signal by quantifying the amount of additive noise in the signal \cite{Joana2018} \cite{ferrand2002}. 
\\

{\bf Pitch}\\
Pitch can be described as the fundamental frequency of the vibrations produced by the vocal cords \cite{mario2016}. This is closely related to frequency because as the frequency increases, so does the pitch. However, they are not exactly equivalent. Pitch is subjective to human perception while frequency is derived based on measurements. Pitch is also known as the human's perception of frequency.
%This is the first appendix.

\subsection{MFCC}
{\bf Pre-Emphasis}\\
This process involves increasing the energy of the signal at higher frequency. This is achieved by passing the speech signal through a filter which emphasizes higher frequency.  This is done because the spectrum for voiced segments has more energy at lower frequency than higher frequency. The aim is to compensate the frequency part that was suppressed during voice production. It achieves this by passing the input speech signal through a filter that amplifies the higher frequencies \cite{parwinder2014}. The output signal can be represented as follows \cite{chavan_2016} \cite{fayek2016}:

\[Y _{n} = X_{n} - 0.95 \times X _{n-1}\]

where: \\
\[Y_{n} = output\,signal\] 
\[X_{n} = input\,signal\]

{\bf Framing}
This process involves splitting the speech signals into frames with each frame containing N sample points \cite{rao_2017}. The width of each frame is usually about 30ms - 40ms with an overlap of half the size of the frame. The duration of a frame can be calculated as follows:

\[d _{f} = \frac{1}{S_{r}} \times k\]

where: \\
\[S_{r} = sampling \, rate\]
\[k = frame \, size \]

{\bf Windowing}
This process involves smoothing signals in order to prevent discontinuity in the frame. This is achieved by multiplying the input signal with a window function. 

A window function is a function used to reduce signal discontinuities at the beginning and end of each frame. This is done by taking the next frame into consideration and integrating the frequency lines, hence making each frame connect smoothly with each other. Hamming window function is the most common window function used for speech analysis. It is described below:

\[W_{n} = 0.54 - 0.46cos(\frac{2\pi n}{N-1})\]\\

where: \\

\[N = number \, of \, samples \, in \, each \, frame\] \\
\[W_{n} = Hamming \, Window\] \\

\[Y _{n} = X_{n} \times W _{n}\]

where: \\
\[Y_{n} = output\,signal\] 
\[X_{n} = input\,signal\]
\[W_{n} = Hamming \, Window\]

{\bf Fast Fourier Transform (FFT)} \label{fft}
%Signals are studied in three main domains.
%\begin{enumerate}[label=(\roman*)]
%\item Frequency Domain: This is the domain for analysis of signals with respect to frequency.
%\item Time Domain: This is the domain for analysis of signals with respect to time.
%\item Wavelet Domain 
%\end{enumerate}

This step involves using Fourier Transform to convert each frame (derived from the previous step) from the time domain to frequency domain. Discrete Fourier Transform (DFT) is an algorithm used to carry out this conversion. Fast Fourier Transform (FFT) is an optimized implementation of DFT.

\[X_k = \displaystyle \sum _{n=0}^{N-1}x_n e^\frac{-j2\pi kn}{N}\]

where:
\[x_n = windowed \, signal\]
\[N = number \, of \, discrete \, frequency \, bands\]
\[X_k = output \, signal \, in \, frequency \, domain\]

{\bf Mel Filter Bank} \label{mel}
The human perception of frequency is non-linear. 

At frequencies below 1 kHz, the mel scale is linear, that is, 1kHz = 1000 mel). At frequencies above 1 kHz, the mel scale is logarithmic. 

The actual frequency needs to be converted to the perceived pitch, measured in mel scale, to mimic the behaviour of the human ear \cite{huang_2001}. The relationship between frequency (Hz) and Mel frequency can be shown as \cite{rao_2017}:

\[f_{mel} = 2595log_{10} (1 + \frac{f}{700})\]

This step involves multiplying the magnitude spectrum, with a set of triangular Mel spaced filter banks to get the Mel spectrum. 

Each filter in the filter bank is triangular and has a response of 1 at the center frequency and decreases until it reaches the center frequencies of the neighboring filters where the response is 0.

Each filter in the filter bank is multiplied by the magnitude spectrum derived previously. The output is known as the Mel spectrum.

\[S_m = \sum _{k=0}^{N}X_k M_{l(k)} \;\]

where:

\[L = total\,number\,of\,triangular\,Mel\,weighting\,filters\]
\[S_{m} = Mel\,spectrum\]
\[X_{k} = initial\,magnitude\,spectrum\]

{\bf Log Energy Computation}\label{log_compute}
The log of the square of the output of the Mel filter bank derived previously is calculated.
\[Y_{m}^\prime = log(\lvert S_{m}^2 \rvert)\]

where:
\[Y_{m}^\prime = Log\,spetral\,vector\]

{\bf Discrete Cosine Transform (DCT)}
This step involves converting the log Mel spectrum to a time-like domain using DCT. This is achieved by applying the DCT formula to the log Mel spectrum output derived in Section \ref{log_compute}. 

\[y_{t(k)} = \displaystyle \sum _{m=1}^{M} log(\lvert Y_{t(m)} \rvert) cos(k(m-0.5) \frac{\pi}{M}\]

where:

\[k = 0,1,2.......L \]
\[L = mel-scale \, ceptral \, coefficients\]
\[M = number \, of \, triangular \, band \, pass \, filters\]

The output is known as  the Mel frequency cepstral coefficients. Most times, about 13 cepstral coefficients from the output are used.

\end{document}